\documentclass[12pt]{article}
\usepackage{float}\restylefloat{figure}
\usepackage{epsfig}
\usepackage{graphicx}
\usepackage{epstopdf}
\def\bm#1{\mbox{\boldmath $#1$}}
\def\beq{\begin{equation}}
\def\eeq{\end{equation}}
\def\t2{\mbox{  }}
\def\rst1{\mbox{ }}

\begin{document} 
\title{Ewald sums for Yukawa potentials in quasi-two-dimensional systems.}
\author{\normalsize Martial MAZARS\footnote{Electronic mail: Martial.Mazars@th.u-psud.fr} \\
\small Laboratoire de Physique Th\'eorique (UMR 8627),\\
\small Universit\'e de Paris XI, B\^atiment 210, 91405 Orsay Cedex,
FRANCE}
\maketitle
LPT 06-98
\begin{center}{\bf Abstract}\end{center}
In this note, we derive Ewald sums for Yukawa potential for three dimensional systems with two dimensional periodicity. This sums are derived from the Ewald sums for Yukawa potentials with three dimensional periodicity  [G. Salin and J.-M. Caillol, J. Chem. Phys. {\bf 113}, 10459 (2000)]  by using the method proposed by Parry for the Coulomb interactions [D.E. Parry, Surf. Sci. {\bf 49}, 433 (1975); {\bf 54}, 195 (1976)].

\newpage
The Yukawa interaction energy between two particles is given by
\begin{equation}
E(r)=\frac{y_i y_j}{\epsilon}\frac{\exp(-\kappa r)}{r}
\end{equation}
where $\epsilon$ is the dielectric constant, $\kappa$ the inverse of the screening length and $y_i$ the "Yukawa charges" defined by the properties and the state of the system ; for instance, at the Debye-H\"uckel approximation for electrolytes or in the Derjaguin-Landau-Verwey-Overbeek (DLVO) theory of colloids, $\kappa$ and $y_i$ are related to physical parameters of systems as
\begin{center}
$\displaystyle \kappa = \sqrt{\frac{q^2\rho}{k_{B}T\epsilon}}\mbox{    }$ and $\mbox{    }\displaystyle y_i=\frac{q\exp(\kappa\sigma_i)}{(1+\kappa\sigma_i)}$
\end{center}
where $\sigma_i$ is the diameter of the hard core of the ions, in the Debye-H\"uckel approximations, or the radius of macroions, in DLVO theory, and $\rho$ and $q$ are respectively the density of ions or counterions and their charge, $k_B$ the Boltzmann constant and $T$ the temperature.\\
Yukawa interactions between particles are used in numerical simulations as effective potentials to simulate systems as plasmas, dusty plasma, colloids, etc. ; on general ground, such potentials may be used as a reasonable approximation, as soon as some microscopic degrees of freedom may be approximated to a continuous background leading to a screening of the direct interaction between particles, while the spherical symmetry of the interaction is preserved.\\
As outlined in ref.[1], if $\kappa$ is large enough, the screening length can be much smaller than simulation box lengths, then interactions between particles are not long ranged and, in practice, a simple truncation of the potential, with the use of the minimum image convention, could be sufficient. On the contrary, if $\kappa$ is not large or quite small, then interactions between particles may be long ranged and images of particles introduced by the periodic boundary conditions may contribute significantly to the energy of the system. In these cases, a crude truncation of the potential could lead to strong bias in computations (for Coulomb interactions, see for instance refs.[2-4] for errors introduced by  crude truncations of long ranged potentials). To handle these latter cases, an Ewald method for systems with three dimensional periodicity and Yukawa interaction potentials has been exhibited$^1$.\\
Many interesting systems which interaction between particles can be approximated by Yukawa potentials are also confined to quasi-two dimensional geometries$^{5,6,7}$, therefore an Ewald method is of interest to permit to simulate the properties of these quasi-two dimensional systems for any value of the $\kappa$ parameter including at low counterions concentration or high temperatures.\\
In this note, we derive Ewald sums for Yukawa potential in quasi-two dimensional systems  from results of ref.[1] following the same derivation done by Parry$^8$ for Coulomb interactions. For Coulomb interaction in quasi-two dimensional systems several methods exist$^{9,10}$, in particular some methods used the Ewald method for three dimensional systems with a highly asymmetric box$^{11,12}$ and by adding correction terms related to the total dipole of the simulation box ; a general review on Coulomb interaction in quasi two-dimensional systems is done in ref.[10]. In a forthcoming work, some numerical implementations on a test system will be given ; the present work is devoted only to provide a simple derivation of Ewald sums for Yukawa potential in quasi-two dimensional systems.\\
As computed by Salin and Caillol$^1$, the Ewald-Yukawa interaction energy is given by 
\begin{equation}
E=E_{\bm{r}}+E_{\bm{k}}-E_{\mbox{\small Self}}
\end{equation}
with the short ranged contribution  
\begin{equation}
\displaystyle E_{\bm{r}}=\frac{1}{4}\sum_{ij}\sum_{\bm{n}}' y_i y_j\frac{D(r_{ij},\bm{n},\kappa ; \alpha)}{\mid \bm{r}_{ij}+\bm{nL}\mid}
\end{equation}
where 
\begin{equation}
\begin{array}{l}
\displaystyle D(r_{ij},\bm{n},\kappa ; \alpha)=\mbox{erfc}(\alpha\mid\bm{r}_{ij}+\bm{nL}\mid+\frac{\kappa}{2\alpha})\exp(\kappa\mid \bm{r}_{ij}+\bm{nL}\mid)\\[0.05in]
\displaystyle +\mbox{erfc}(\alpha\mid \bm{r}_{ij}+\bm{nL}\mid-\frac{\kappa}{2\alpha})\exp(-\kappa\mid \bm{r}_{ij}+\bm{nL}\mid)
\end{array}
\end{equation}
and $r_{ij}$ the distance between the pair $(i,j)$ of particles, the long ranged contribution 
\begin{equation}
E_{\bm{k}} = \frac{2\pi}{V}\sum_{\bm{k}\neq 0}\frac{\exp(-\bm{k}^2+\kappa^2)/4\alpha^2)}{\bm{k}^2+\kappa^2}\mbox{\large{$\mid$}}\sum_{i} y_i\exp(i\bm{k}\mbox{.}\bm{r}_i)\mbox{\large{$\mid$}}^2
\end{equation}
and the self interaction 
\begin{equation}
\displaystyle E_{\mbox{\small Self}}=\mbox{\Huge{[}} \frac{\alpha}{\sqrt{\pi}}\exp\mbox{\large{(}}-\frac{\kappa^2}{4\alpha^2}\mbox{\large{)}}-\frac{\kappa}{2}\mbox{ }\mbox{erfc}\mbox{\large{(}}\frac{\kappa}{2\alpha}\mbox{\large{)}}\mbox{\Huge{]}}\sum_i y_i^2
\end{equation}
where we have set $\epsilon=1$ and used conventional notations for Ewald sums ; namely, $V$ is the volume of the simulation box, $\bm{nL}$ the condensed notation for the vectors of the periodic boundary conditions, $\bm{k}$ are the vectors belonging to the reciprocal lattice associated with the three dimensional periodicity and $\alpha$ the damping parameter of the Ewald method. In Eq.(3), the prime in the sum over $\bm{n}$ indicates that for $\bm{n}=0$, the self terms $i=j$ are not included. The short ranged contributions for the Ewald sums  for Yukawa potentials in periodic three dimensional systems are given as in Eq.(3), for quasi-two dimensional$^8$ the short ranged contributions are also given by Eq.(3) ; for pratical applications, a choice of the Ewald damping parameter $\alpha$ is done such that  summations over images are restricted to the minimum image convention$^1$.\\
In the work by Parry$^8$, the Ewald sums for quasi-two dimensional systems are derived from the Ewald method for three dimensional systems by letting the spatial periodicity along the third direction to tend to infinity ($L_z\rightarrow\infty$).\\ 
In the following, we use the notations $\bm{k}=\bm{G}+k\bm{e}_z$ where $\bm{G}$ are the vectors belonging to the reciprocal lattice associated with the two dimensional periodicity and $k=2\pi m/L_z$, ($m$ integer), where $L_z$ is the spatial periodicity of the simulation box along the $\bm{e}_z$ ; we set also $\bm{r}_{ij}=\bm{s}_{ij}+z_{ij}\bm{e}_z$ and $V=AL_z$, where $s_{ij}$ is the distance in the plane perpendicular to $\bm{e}_z$ and $A$ is the surface of the simulation box for quasi-two dimensional systems.\\
Following Parry$^8$, the long ranged contribution is separated into two contributions. The first contribution$^{8(a)}$, noted $E_{G\neq 0}^{(a)}$, is obtained for $\bm{G}\neq 0$ and the second$^{8(b)}$ for $\bm{G} = 0$, as a reminiscent contribution of summation over $k$ as $L_z\rightarrow\infty$, noted below $E_{G= 0}^{(b)}$.\\
From Eq.(5), we have 
\begin{equation}
\begin{array}{c}
\displaystyle E_{G\neq 0}^{(a)}=\frac{2\pi}{A}\sum_{ij} y_iy_j\sum_{\bm{G}\neq 0}\exp(i\bm{G}\mbox{.}\bm{s}_{ij})\exp(-(\bm{G}^2+\kappa^2)/4\alpha^2)\\[0.05in]
\displaystyle \times\lim_{L_z\rightarrow\infty}\frac{1}{L_z}\sum_{r=-\infty}^{+\infty}\frac{\exp(-k^2/4\alpha^2)}{k^2+(\bm{G}^2+\kappa^2)}\exp(ikz_{ij})
\end{array}
\end{equation}
Taking into account the identity$^{13}$ 
\begin{equation}
\begin{array}{l}
\displaystyle\lim_{L\rightarrow\infty}\frac{1}{L}\sum_{r=-\infty}^{+\infty}\frac{\exp(-k^2/4\alpha^2)}{k^2+(\bm{G}^2+\kappa^2)}\exp(ikz_{ij})=\frac{1}{2\pi}\int_{-\infty}^{+\infty}dk\frac{\exp(ikz)\exp(-k^2/4\alpha^2)}{k^2+(\bm{G}^2+\kappa^2)} 
\end{array}
\end{equation}
we found 
\begin{equation}
\displaystyle E_{G\neq 0}^{(a)}=\frac{\pi}{2A}\sum_{\bm{G}\neq 0}\sum_{ij} y_iy_j F(\sqrt{\bm{G}^2+\kappa^2},z_{ij};\alpha) \exp(i\bm{G}\mbox{.}\bm{s}_{ij})
\end{equation}
with
\begin{equation}
\begin{array}{c}
\displaystyle F(K,z;\alpha)=\frac{1}{K}\mbox{\Large{(}}\exp(K z)\mbox{erfc}\mbox{\large{(}}\frac{K}{2\alpha}+\alpha z\mbox{\large{)}}+\exp(-K z)\mbox{erfc}\mbox{\large{(}}\frac{K}{2\alpha}-\alpha z\mbox{\large{)}}\mbox{\Large{)}}
\end{array}
\end{equation}
The second contribution is given by
\begin{equation}
E_{G= 0}^{(b)}=\frac{\pi}{2A}\sum_{ij} y_iy_j F(\kappa,z_{ij};\alpha)
\end{equation}
As it is shown explicitly on Eqs.(9) and (11), the long ranged contributions of Ewald-Yukawa sums for quasi-two dimensional systems can not be reduced  to one particle summation, because of the complicated dependence on $z_{ij}$.\\
Self energy is given by 
\begin{equation}
\left \{\begin{array}{ll}
\displaystyle E^{Q2D}(\mbox{\small Self})&\displaystyle = \mbox{\Huge{[}}\frac{\alpha}{\sqrt{\pi}}\exp(-\kappa^2/4\alpha^2)-\frac{\kappa}{2}\mbox{ erfc}(\kappa/2\alpha)\mbox{\Huge{]}}\sum_i y_i^2 \\
&\\
\displaystyle E_{G= 0}(\mbox{\small Self})&\displaystyle = \frac{\pi}{A}\frac{\mbox{erfc}(\kappa/2\alpha)}{\kappa}\sum_i y_i^2
\end{array}
\right.
\end{equation}
The Ewald sums for Yukawa potential in quasi-two dimensional systems are given by
\begin{equation}
E=E_{\bm{r}}+E_{G\neq 0}^{(a)}+E_{G= 0}^{(b)}-E^{Q2D}(\mbox{\small Self})-E_{G= 0}(\mbox{\small Self})
\end{equation}
with each contribution given by Eqs.(3,9,11) and (12).\\
When $\kappa\rightarrow 0$, the Helmholtz equation, which solutions lead to Yukawa potentials, becomes a Poisson equation, which solutions lead to Coulomb potentials$^1$ ; then the Ewald sums for Coulomb interactions may be recovered from Eqs.(3,9,11) and (12), under the condition that the electroneutrality of the periodic system is recovered too$^{14,15}$ (for instance by letting the counterions concentration acting as an uniform continuous background).\\ 
When $z_{ij}=0$ for all pairs of particles, then all particles are confined in a plan. In this limit, the Ewald sums for a quasi-two dimensional systems reduce to the Ewald sums of  two dimensional systems. Then, Eqs.(9,12) and (13) become 
\begin{equation}
E^{2D}=E_{\bm{s}}+E_{G\neq 0}-E^{2D}_{\mbox{\small Self}}
\end{equation}
with
\begin{equation}
\left \{\begin{array}{ll}
\displaystyle E_{G\neq 0}&\displaystyle =\frac{\pi}{A}\sum_{\bm{G}\neq 0}\frac{\mbox{erfc}(\sqrt{\bm{G}^2+\kappa^2}/2\alpha)}{\sqrt{\bm{G}^2+\kappa^2}}\mbox{\large{$\mid$}}\sum_{i} y_i \exp(i\bm{G}\mbox{.}\bm{s}_{i})\mbox{\large{$\mid$}}^2\\
&\\
\displaystyle E^{2D}_{\mbox{\small Self}}&\displaystyle = \mbox{\Huge{[}}\frac{\alpha}{\sqrt{\pi}}\exp(-\kappa^2/4\alpha^2)-\frac{\kappa}{2}\mbox{ erfc}(\kappa/2\alpha)\mbox{\Huge{]}}\sum_i y_i^2
\end{array}
\right.
\end{equation}
and $E_{\bm{s}}$ given by Eq.(3), with $\bm{s}_{ij}$ instead of $\bm{r}_{ij}$.\\
These results may be applied directly in computer simulations, nevertheless the efficiency suffers from the lack of reduction in one particle summations, as for Coulomb potentials in quasi-two dimensional systems, because of the complicated dependence on $z_{ij}$. For molecular dynamics implementations, the forces acting on particles can easily be obtained with computations of the derivatives of the interaction potential.

\newpage
\vspace{.5cm}
{\bf\Large References}\\[0.1in]
\normalsize 
$^1$ G. Salin and J.-M. Caillol, J. Chem. Phys., {\bf 113}, 10459 (2000)\\
$^2$ J.C. Shelley and G.N. Patey,  Mol. Phys., {\bf 88}, 385 (1996)\\
$^3$ M. Patra, M. Karttunen, M.T. Hyv\"{o}nen, E. Falck, P. Lindqvist and I. Vattulainen,  Biophys. J., {\bf 84}, 3636 (2003)\\
$^4$ M. Bergdorf, C. Peter and P.H. H\"onenberger, J. Chem. Phys., {\bf 119}, 9129 (2003)\\
$^5$ S.H. Behrens and D.G. Grier, Phys. Rev. E, {\bf 64}, 050401 (2001)\\
$^6$ J. Santana-Solano and J.L. Arauz-Lara, Phys. Rev. E, {\bf 65}, 021406 (2002)\\
$^7$ Y. Han and D.G. Grier, J. Chem. Phys. {\bf 122}, 064907 (2005)\\
$^8$ (a) D.E. Parry, Surf. Sci. {\bf 49}, 433 (1975) ; (b) {\bf 54}, 195 (1976)\\
$^9$ D. M. Heyes, M. Barber and J. H. R. Clarke,  J. Chem. Soc. Faraday Trans. 2, {\bf 73}, 1485 (1977)\\
$^{10}$ M. Mazars, Mol. Phys., {\bf 103}, 1241 (2005)\\ 
$^{11}$ E. Spohr,  J. Chem. Phys., {\bf 107}, 6342 (1997)\\
$^{12}$ I.-C. Yeh and M.L. Berkowitz, J. Chem. Phys., {\bf 111}, 3155 (1999)\\ 
$^{13}$ M. Mazars, J. Chem. Phys., {\bf 117}, 3524 (2002)\\
$^{14}$ A. Grzybowski, E. Gw\'o\'zd\'z,  and A. Br\'odka, Phys. Rev. B,  {\bf 61}, 6706 (2001)\\
$^{15}$ An analysis of the diverging behavior as $\kappa\rightarrow 0$ is done in : M. Mazars, Mol. Phys. {\bf 103}, 675 (2005) ; (cf. Eqs.(2) and (4) of the cited work, with $\beta\equiv\kappa$ ; it is worthwhile to note also that Eq.(3) provides the Lekner sums for Yukawa potentials). Lekner sums for Coulomb interactions have been defined in J. Lekner,  Physica A, {\bf 157}, 826 (1989) ; for numerical implementations of this method see for instance : M.Mazars, J. Chem. Phys., {\bf 115}, 2955 (2001) and ref.[10]. 

\end{document}